\documentstyle[12pt]{article}
\begin{document}
\hfill{NCKU-HEP-98-01/NHCU-HEP-98-01}
\vskip 0.3cm
\begin{center}
{\large {\bf Origin of the $k_T$ smearing in direct photon production}}
\vskip 1.0cm
Hung-Liang Lai
\vskip 0.3cm
Department of Physics, National Tsing-Hua University, \par
Hsinchu 300, Taiwan
\vskip 0.5cm
Hsiang-nan Li
\vskip 0.3cm
Department of Physics, National Cheng-Kung University, \par
Tainan 701, Taiwan
\end{center}
\vskip 1.0cm

PACS numbers: 12.38.Bx, 12.38.Cy, 13.60.Hb
\vskip 1.0cm

\centerline{\bf Abstract}
\vskip 0.3cm
We show that the Sudakov factor from the resummation of double logarithms
$\ln(s/k_T^2)$ contained in the distribution functions is
responsible for the
$k_T$ smearing mechanism employed in the next-to-leading-order QCD
($\alpha\alpha_s^2$) calculations of direct photon production. $s$ is the
center-of-mass energy, and $k_T$ the transverse momentum carried by a
parton in a colliding hadron. This factor exhibits the appropriate
$s$-dependent Gaussian width in $k_T$, such that our predictions are in
good agreement with experimental data.

\newpage

\centerline{\large \bf I. INTRODUCTION}
\vskip 0.5cm

It was pointed out some time ago \cite{HKK} that the global QCD analysis  
of the direct photon production
processes from both fixed-target and collider experiments
\cite{WA70,UA6B,E706,UA6S,R806,UA2,CDF1,CDF2} has led to a
puzzle: Most data sets show a steeper $p_T$ distribution than the
next-to-leading-order (NLO) QCD ($\alpha\alpha_s^2$)
predictions, $p_T$ being the transverse momentum of the direct photon.
This behavior can not be explained by either global fits with new parton
distribution functions or improved photon fragmentation functions. To
resolve the puzzle, a Gaussian type broadening of the
transverse momentum $k_T$ carried by initial-state partons in colliding
hadrons has been introduced \cite{HKK}.
This smearing function enhances the low end region of the spectrum in $p_T$
more than the high end, such that the QCD predictions have a steeper
distribution. The recent publication of the E706 direct photon data set 
\cite{E706_97} has confirmed the evidence for the $k_T$ effect. However, 
the physical origin of this effect and
of its center-of-mass-energy dependent Gaussian width has not been
understood yet.

In this paper we shall propose the mechanism that is responsible for the
$k_T$-smearing effect in direct photon production. It has been observed
that large double logarithms $\ln^2(s/k_T^2)$ are generated, $s$ being the
center-of-mass energy, when the transverse degrees of freedom of the
partons are taken into account \cite{L1}. The large logarithms are
absorbed into parton distribution functions, and their all-order summation
leads to a Sudakov factor, which describes the perturbative distribution
of the partons in $k_T$ for different $s$. Since the Sudakov factor gives
strong suppression at large $k_T$, it resembles a Gaussian function with
$s$-dependent width. These characteristics are qualitatively consistent
with those of the smearing effect mentioned above. We shall demonstrate
quantitatively that the inclusion of the Sudakov factor indeed modifies
the NLO QCD ($\alpha\alpha_s^2$) predictions in
the desired way, and the data of direct photon production can be explained.

Due to the transverse degrees of freedom of the partons, the resummation
of double logarithms should be performed in the impact parameter $b$ space
\cite{L1,CS}, which is conjugate to $k_T$. Hence, it is $\ln^2(s b^2)$ that
are resummed. After deriving the Sudakov factor, we Fourier transform it
back to the $k_T$ space, and convolute it with the naive NLO factorization
formula for direct photon production. Note that an analytical expression
of the transformed Sudakov factor does not exist, and must be obtained
numerically. However, to make its smearing effect transparent, the Sudakov
factor is parametrized by a Gaussian function $\exp(-\Gamma^2 b^2/4)$
with the width $\Gamma$, which corresponds to $\exp(-k_T^2/\Gamma^2)$ in
the $k_T$ space. We find that the $s$ dependence of $\Gamma$ is consistent
with that employed in \cite{HKK}, implying the success of our analysis.

In Sect. II we apply the Collins-Soper-Sterman resummation technique
\cite{CS} to direct photon production from hadron collisions, and
present the explicit expression of the Sudakov factor. In Sect. III the
Sudakov factor is parametrized as a Gaussian function, and the $s$
dependence of its width is examined. The numerical results are
compared with the data of direct photon production. Section IV is
the conclusion.
\vskip 1.0cm

\centerline{\large \bf II. FACTORIZATION AND RESUMMATION}
\vskip 0.5cm
                                                
In this section we derive the factorization formula for direct photon
production
\begin{equation}
h(p_1)+h(p_2) \to \gamma(p_T) + X\;,
\label{dy}
\end{equation}
in the collision of the hadrons $h$, and resum the involved large double
logarithms into a Sudakov factor. The hadron momenta are assigned as
$p_1=(p_1^+,0,{\bf 0}_T)$ and $p_2=(0,p_2^-,{\bf 0}_T)$ with $p_1^+= 
p_2^-=\sqrt{s/2}$. $p_T$ is the
transverse momentum of the direct photon that is measured. The tree-level
diagrams are shown in Figs.~1(a) and 1(b), in which both quarks and gluons
out of the hadrons contribute. The partons carry the momenta
$\xi_i p_i+{\bf k}_{iT}$, where $\xi_i$, $i=1$, 2, are the
longitudinal momentum fractions and ${\bf k}_{iT}$ are the
transverse momenta. ${\bf k}_{iT}$ flow into the direct photon, and make up
the momentum $p_T$.

We then consider higher-order corrections to Fig.~1(a) 
from Figs.~1(c)-1(h). The discussion for Fig.~1(b) is similar. Fig.~1(c),
the self-energy correction to a parton, and Fig.~1(d), the loop correction
with a real gluon connecting the two partons from the same hadron, contain
both collinear divergences from the loop momentum $l$ parallel to $p_i$ and
soft divergences from small $l$. Since the soft divergences cancel
asymptotically as shown in \cite{L1}, the double logarithmic corrections
are mainly collinear. Therefore, Figs.~1(c) and 1(d) are absorbed into a
distribution function $\phi$ associated with the hadron $i$. In most of
kinematic regions the self-energy correction to the outgoing jet in
Fig.~1(e) does not give collinear divergences. Because of the cancellation
of soft divergences, Fig.~1(e) is absorbed into a hard scattering amplitude
$H$, which corresponds to the parton-level differential cross section
$d\sigma/dp_T$. Similarly, other radiative corrections to the box diagram
are also infrared finite and grouped into $H$.

The absorption of the irreducible diagram in Fig.~1(f) is more
delicate. It contains collinear divergences from $l$ parallel to 
$p_i$ and soft divergences. For $l$ parallel to $p_1$, we replace the
parton line in hadron 2 by an eikonal line in the direction
$n=v_2=p_2/\sqrt{s/2}$ on the light cone, and factorize the gluon into
the distribution function of hadron 1 from the full scattering amplitude.
The replacement by an eikonal line holds for both quarks and gluons as
verified in \cite{L2}. The treatment of the gluon with $l$ parallel to
$p_2$ is the same, but the direction of the eikonal line is
$n=v_1=p_1/\sqrt{s/2}$. The factorization of the vertex-correction
diagrams in Figs.~1(g) and 1(h) is similar to that of Fig.~1(f). In the
leading region with the loop momentum $l$ parallel to $p_i$, they are
assigned to the corresponding distribution functions. Certainly, soft
divergences in the above diagrams cancel asymptotically. At last, when $l$
is hard, Figs.~1(c)-1(h) are grouped into the hard scattering amplitude
$H$ defined above.

The eikonal lines on the light cone introduced above, collecting collinear
gluons, are essential for the factorization of irreducible diagrams in
Figs.~1(e)-1(h), and for the gauge invariance of the parton distribution
functions. However, to implement the resummation technique, we allow
the gauge vector $n$ to vary away from the light cone ($n^2\not= 0$) 
temporarily. It will
be shown that the Sudakov factor turns out to be $n$-independent. After
completing the resummation, $n$ is brought back to the light cone, and the
gauge invariance of the parton distribution functions is restored. That
is, the arbitrary $n$ appears only at the intermediate stage of the
formalism and as an auxiliary tool of the resummation.

When the transverse degrees of freedom of the partons are taken
into account, the factorization must be performed in the 
$b$ space, which is conjugate to $k_T$ \cite{L1}.
Hence, we arrive at the general factorization picture for direct photon
production shown in Fig.~2, and the corresponding formula,
\begin{eqnarray}
\frac{d\sigma(s,p_T)}{dp_T}&=&
\int d\xi_1 d\xi_2 \int \frac{d^2{\bf b}}{(2\pi)^2}
{\tilde \phi}(\xi_1,p_1,b,\mu){\tilde \phi}(\xi_2,p_2,b,\mu)
\nonumber \\
& &\times {\tilde H}(\xi_1,\xi_2,s,b,\mu)\exp(i{\bf p}_T\cdot{\bf b})\;,
\label{fdy}
\end{eqnarray}
$\mu$ being a renormalization and factorization scale. We have assumed
the single $b$ dependence in the above expression in order to compare our
formalism with the naive $k_T$ smearing employed in the literature. A
rigorous factorization formula for direct photon production from hadron
collisions will be presented elsewhere.

In the axial gauge $n\cdot A=0$ the eikonal line disappears, and only
Figs.~1(c) and 1(d) give double logarithmic corrections to the distribution
functions \cite{L1}. We demonstrate how to resum the double logarithms
from Figs.~1(c) and 1(d) by considering only the hadron 1. The resummation
for the hadron 2 is the same. The essential step in the resummation is to
derive a differential equation $p_1^+d{\tilde \phi}/dp_1^+=C{\tilde \phi}$
\cite{L1}, where the coefficient function $C$ contains only single
logarithms, and can be treated by RG methods. In the axial gauge $n$ goes
into the gluon propagator, $(-i/l^2)N^{\mu\nu}(l)$, with
\begin{equation}
N^{\mu\nu}(l)=g^{\mu\nu}-\frac{n^\mu l^\nu+n^\nu l^\mu}
{n\cdot l}+n^2\frac{l^\mu l^\nu}{(n\cdot l)^2}\;.
\label{gp}
\end{equation}
Because of the scale invariance of $N^{\mu\nu}$ in $n$, ${\tilde \phi}$
must depend on $p_1$ (or $s$) through the ratio $(p_1\cdot n)^2/n^2$,
implying that 
the differential operator $d/dp_1^+$ can be replaced by $d/dn$ using a
chain rule,
\begin{equation}
p_1^+\frac{d}{dp_1^+}{\tilde \phi}=-\frac{n^2}{v_1\cdot n}v_{1\alpha}
\frac{d}{dn_\alpha}{\tilde \phi}\;.
\label{cr}
\end{equation}

The operator $d/dn_\alpha$ applying to $N^{\mu\nu}$ gives
\begin{eqnarray}
-\frac{n^2}{v_1\cdot n}v_{1\alpha}
\frac{d}{dn_\alpha}N^{\mu\nu}= {\hat v}_{\alpha}
\left(N^{\mu\alpha}l^\nu+N^{\alpha\nu}l^\mu\right)\;,
\label{dgp}
\end{eqnarray}
with the special vertex
\begin{equation}
{\hat v}_{\alpha}=\frac{n^2v_{1\alpha}}{v_1\cdot nn\cdot l}\;.
\end{equation}
The momentum $l^\mu$ ($l^\nu$) appearing at the end of the differentiated
gluon line is contracted with a vertex the gluon attaches, which is then
replaced by the special vertex. The contraction of $l^\mu$ hints the
application of the Ward identity. Summing all the diagrams with different
differentiated gluons, the special vertex moves to the outer
end of the parton line due to the Ward identity. We obtain the derivative,
\begin{equation}
p_1^+\frac{d}{dp_1^+}{\tilde \phi}=2{\tilde \phi}'\;,
\end{equation}
described by Fig.~3(a), where the square in the new function
${\tilde \phi}'$ represents the special vertex ${\hat v}_\alpha$. The
coefficient 2 comes from the equality of the two new functions with
the special vertex on either of the two parton lines. 

The collinear region of the loop momentum $l$ is now not important because
of the factor $1/(n\cdot l)$ in ${\hat v}_\alpha$ with nonvanishing $n^2$.
Therefore, the leading regions of $l$ are soft and hard, in which the
subdiagram containing the special vertex can be factorized from
${\tilde \phi}'$ into a function $K$ and a function $G$, respectively. The
remaining part is the original distribution function ${\tilde \phi}$. The
differential equation is then expressed as
\begin{equation}
p_1^+\frac{d}{dp_1^+}{\tilde \phi}=2\left[K(b\mu,\alpha_s(\mu))+
G(p_1^+/\mu,\alpha_s(\mu))\right]{\tilde \phi}\;,
\label{dph}
\end{equation}
with the general diagrams of $K$ and $G$ shown in Fig.~3(b). The sum
$K+G$ is exactly the coefficient function $C$ mentioned above. It has
been made explicit that $K$ contains the single small scale $1/b$ and $G$
contains the single large scale $p_1^+$.

The $O(\alpha_s)$ contributions to $K$ from Fig.~3(c) and to $G$ from 
Fig.~3(d) are given by
\begin{eqnarray}
K&=&ig^2{\cal C}\mu^\epsilon\int\frac{d^{4-\epsilon}l}
{(2\pi)^{4-\epsilon}}
\left[\frac{1}{l^2}+2\pi i\delta(l^2)e^{i{\bf l}_T\cdot {\bf b}}\right]
\frac{{\hat v}_\mu v_\nu}{v\cdot l}N^{\mu\nu}-\delta K\;,
\label{kph}\\
G&=&ig^2{\cal C}\mu^\epsilon\int\frac{d^{4-\epsilon}l}
{(2\pi)^{4-\epsilon}}{\hat v}_\mu
\left(\frac{\not p+\not l}{(p+l)^2}\gamma_\nu
-\frac{v_\nu}{v\cdot l}\right)\frac{N^{\mu\nu}}{l^2}-\delta G\;,
\label{gph}
\end{eqnarray}
$\delta K$ and $\delta G$ being the corresponding additive counterterms.
The color factor ${\cal C}$ is $C_F (=4/3)$ if the parton is a quark,
and $N_c (=3)$ if the parton is a gluon \cite{L2}. The factor
$e^{i{\bf l}_T\cdot {\bf b}}$ in Eq.~(\ref{kph}), which is associated with
the second diagram (real gluon emission) in Fig.~3(d) \cite{L1,CS}, renders
$K$ free of infrared poles. That is, $1/b$ serves as an infrared cutoff of
the loop integral. Note that $K$ vanishes in the asymptotic region
$b\to 0$, reflecting the soft cancellation as stated before. The second
term in Eq.~(\ref{gph}), as a soft subtraction, ensures that the
momentum flow in $G$ is hard.

A straightforward calculation of Eqs.~(\ref{kph}) and (\ref{gph})
gives the pole terms $\delta K=-\delta G$. For details of the calculation,
refer to \cite{L1}. Since $K$ and $G$ contain only single soft and
ultraviolet logarithms, respectively, they are treated by RG methods:
\begin{equation}
\mu\frac{d}{d\mu}K=-\lambda_K=
-\mu\frac{d}{d\mu}G\;.
\label{kg}
\end{equation}
The anomalous dimension of $K$, $\lambda_K=\mu d\delta K/d\mu$,
is given, up to two loops, by \cite{BS}
\begin{equation}
\lambda_K=\frac{\alpha_s}{\pi}{\cal C}+\left(\frac{\alpha_s}{\pi}
\right)^2{\cal C}\left[{\cal C}_A\left(\frac{67}{36}
-\frac{\pi^{2}}{12}\right)-\frac{5}{18}n_{f}\right]\;,
\label{lk}
\end{equation}
with $n_{f}$ the number of quark flavors, and ${\cal C}_A=3$ a color 
factor. In solving Eq.~(\ref{kg}), we make the scale $\mu$ evolve to 
the infrared cutoff $1/b$ in $K$ and to $p^+$ in $G$. The RG solution of 
$K+G$ is written as
\begin{eqnarray}
K(b\mu,\alpha_s(\mu))+G(p^+/\mu,\alpha_s(\mu))=
-\int_{1/b}^{p^+}\frac{d{\bar\mu}}{\bar\mu}
\lambda_K(\alpha_s({\bar\mu}))\;.
\label{skg}
\end{eqnarray}

Substituting Eq.~(\ref{skg}) into (\ref{dph}), we obtain the solution
\begin{eqnarray}
{\tilde \phi}(\xi_1,p_1,b,\mu)=f(\xi_1 p_1^+,b)
{\tilde \phi}(\xi_1,b,\mu)
&\approx & f(\xi_1 p_1^+,b)\phi(\xi_1,\mu)\;,
\label{sph}
\end{eqnarray}
with the Sudakov factor
\begin{equation}
f(\xi_1 p_1^+,b)=\exp\left[-2\int_{1/b}^{\xi_1 p_1^+}\frac{d p}{p}
\int_{1/b}^{p}\frac{d{\bar \mu}}{\bar \mu}
\lambda_{K}(\alpha_s({\bar \mu}))\right]\;.
\label{fb}
\end{equation}
We have set the lower bound of the variable $p$ to $1/b$, and the upper 
bound to $\xi_1 p_1^+$. Note that Eq.~(\ref{fb}) is defined only for
$\xi_1 p_1^+ \ge 1/b$. For $\xi_1 p_1^+ < 1/b$, we require $f$ to be
equal to unity. We also set $f$ to 1 as $f>1$ \cite{LS}, which occurs
in the small $b$ region. The radiative corrections in this short-distance
region should be absorbed into the hard scattering amplitude $H$,
instead of into the distribution function, giving its Sudakov evolution.
The wave function $\phi(\xi_1,\mu)=\int d^2{\bf k}_{1T}
\phi(\xi_1,k_{1T},\mu)$ is the $b\to 0$ limit of
the initial condition ${\tilde \phi}(\xi_1,b,\mu)$ of the Sudakov
evolution. This approximation is reasonable because of the strong
suppression of $f$ at large $b$. Obviously, $f$ is
independent of the vector $n$. Now we make $n$ approach $v_2$ (the light
cone), and $\phi(\xi_1,\mu)$ coincides with the standard distribution
function with the gauge invariance.

The resummation for the distribution function of hadron 2 gives a similar
Sudakov factor. Hence, the additional smearing factor compared to the
standard NLO QCD calculations is given by,
\begin{equation}
{\tilde S}(\xi_1,\xi_2,s,b)=f(\xi_1 p_1^+,b)f(\xi_2 p_2^-,b)\;.
\label{sff}
\end{equation}
Transformation of Eq.~(\ref{fdy}) back to the $k_T$ space with
Eqs.~(\ref{sph}) and (\ref{sff}) inserted leads to
\begin{eqnarray}
\frac{d\sigma(s,p_T)}{dp_T}&=&
\int d\xi_1 d\xi_2 \int d^2{\bf k}_T
\phi(\xi_1,p'_T)\phi(\xi_2,p'_T)
\nonumber \\
& &\times H(\xi_1,\xi_2,s,p'_T)S(\xi_1,\xi_2,s,k_T)\;,
\label{fdk}
\end{eqnarray}
with the variable $p'_T=|{\bf p}_T-{\bf k}_T|$. We have set $\mu$ of
$\phi$ and $H$ to the characteristic scale $p'_T$ of $H$. That is, the
distribution functions $\phi$ will evolve to $p'_T$ according to the
Dokshitzer-Gribov-Lipatov-Altarelli-Parisi equation \cite{AP}, which
further incorporates the summation of the single logarithms $\ln p'_T$.
Equation (\ref{fdk}) is the factorization formula for direct photon
production with the single $k_T$ approximation.

To be related to the analysis in \cite{HKK}, we shall not work on
Eq.~(\ref{fdk}), but neglect the $\xi$ dependence of $S$, making
the approximation $S(s,k_T)\approx S(0.5,0.5,s,k_T)$. It is reasonable
to assume $\xi_1$, $\xi_2\approx 0.5$. Performing the
integration over $\xi_1$ and $\xi_2$, Eq.~(\ref{fdk}) reduces to
\begin{eqnarray}
\frac{d\sigma(s,p_T)}{dp_T}=
\int d^2{\bf k}_T\frac{d\sigma(s,p'_T)}{dp'_T}S(s,k_T)\;,
\label{fa}
\end{eqnarray}
where the differential cross section
\begin{eqnarray}
\frac{d\sigma(s,p'_T)}{dp'_T}=
\int d\xi_1 d\xi_2 \phi(\xi_1,p'_T)\phi(\xi_2,p'_T)
H(\xi_1,\xi_2,s,p'_T)\;.
\end{eqnarray}
is identified as the standard NLO QCD predictions, and $S(s,k_T)$ as
the smearing function we shall investigate. Note that $S$ for Figs.~1(a)
and 1(b) are different, and thus Eq.~(\ref{fa}) in fact represents the
sum over the two diagrams. While the smearing function employed in
\cite{HKK} is the same for Figs.~1(a) and 1(b).

\vskip 1.0cm

\centerline{\large \bf III. NUMERICAL ANALYSIS}
\vskip 0.5cm

Before proceeding with the numerical analysis of Eq.~(\ref{fa}),
we examine the smearing effect of the Sudakov factor $S$. It is known that
the Gaussian smearing function $\exp(-k_T^2/\Gamma^2)$ employed in
\cite{HKK} for direct photon production possesses the $s$-dependent width
$\Gamma$, which is summarized as
\begin{eqnarray}
& &\Gamma \le 1\;\;{\rm GeV}\;\;\;\;\;\;\;\;\;\;
{\rm for}\;\;\sqrt{s}\sim 30\;\;{\rm GeV}\;,
\nonumber \\
& &\Gamma \sim 1-2\;\;{\rm GeV}\;\;\;\;{\rm for}\;\;\sqrt{s}\sim 30-100
\;\;{\rm GeV}\;,
\nonumber \\
& &\Gamma \sim 2-3\;\;{\rm GeV}\;\;\;\;{\rm for}\;\;\sqrt{s}\sim 100-600
\;\;{\rm GeV}\;,
\nonumber \\
& &\Gamma \sim 3-4\;\;{\rm GeV}\;\;\;\;{\rm for}\;\;\sqrt{s}\sim 600-1800
\;\;{\rm GeV}\;.
\label{gs}
\end{eqnarray}
That is, $\Gamma$ increases with $s$ slightly. The predictions
are very sensitive to the $s$ dependence of $\Gamma$. Hence, it is
nontrivial that the experimental data can be explained by our formalism.

We demonstrate that the factor $S(s,k_T)$ exhibits the desired behavior
shown in Eq.~(\ref{gs}). For simplicity, we parametrize
$f(\xi\sqrt{s/2},b)$ as a Gaussian function $\exp(-\Gamma^2 b^2/4)$ with
$\Gamma^2=c(\xi\sqrt{s})^r$, which is the Fourier transformation of
$\exp(-k_T^2/\Gamma^2)$. The momentum fraction $\xi$ will be set to
1/2 in the numerical analysis below. The constants $c$ and $r$ are
determined from the best fit of the parametrization to $f$, considering the
variation of $b$ between 0 and $1/\Lambda_{\rm QCD}=5$ GeV$^{-1}$, and of
$\sqrt{s}$ between 20 and 1800 GeV. The results are $c=0.12$ and $r=0.60$,
if the parton is a quark, and $c=0.09$ and $r=0.83$, if the parton is a
gluon. The decrease of $f(\sqrt{s/2},b)$ with $b$ for different $\sqrt{s}$
and its corresponding parametrization $\exp(-c\sqrt{s}^r b^2/4)$ from best
fit are shown in Fig.~4. The values $f=1$ at small $b$ come from the
truncation of $f>1$ argued before. It is observed that the $s$ dependence
of the width $\Gamma$ is roughly in agreement with Eq.~(\ref{gs}).

We then compute the differential cross section of direct photon production
based on Eq.~(\ref{fa}) with the parametrized smearing function,
\begin{equation}
S(s,k_T)=\exp[-k_T^2/(\Gamma_1^2+\Gamma_2^2)]\;,
\label{s12}
\end{equation}
where the widths $\Gamma_{1(2)}=c(\sqrt{s}/2)^r$ are associated the
hadron 1(2). We can certainly employ the exact Sudakov factor $f$, and
convert it into the $k_T$ space numerically, when evaluating Eq.~(\ref{fa}).
This approach is not followed here, simply because we intend to make our
analysis in full analogy with that in \cite{HKK}. In Fig.~5 we show the
deviation (Data -Theory)/Theory of the NLO QCD predictions, obtained using
the CTEQ4M parton distributions \cite{cteq4}, from the experimental data as
a function of $x_t=2p_T/\sqrt{s}$. Obviously, the deviation is huge,
especially at low $x_t$ of each set of the data. In Fig.~6 the theoretical
predictions come from Eq.~(\ref{fa}) which includes the $k_T$ smearing in
Eq.~(\ref{s12}). It is clear that a significant improvement on the
agreement between theory and experiments is achieved.

\vskip 1.0cm

\centerline{\large \bf IV. CONCLUSION}
\vskip 0.5cm

In this paper we have identified the $k_T$ smearing, which is essential
for the explanation of the direct photon production data, as the Sudakov
factor from the resummation of large radiative corrections to the parton
distribution functions. The identification is confirmed by examining the
$s$ dependence of the Gaussian widths of the parametrized Sudakov factors,
and the consistency of the improved predictions with the data. Compared to
our analysis, the smearing employed in \cite{HKK} is very naive. The
effects should depend on the type of partons, and on the momentum
fractions. A more accurate analysis even involves the recalculation of the
hard scattering amplitudes with the partons off shell by $k_{1T}$ and
$k_{2T}$, which are associated with the hadron 1 and 2, respectively.
Our formalism including the resummation at low $k_T$ can be generalized to
other QCD processes. All these subjects will be studied elsewhere.
\vskip 1.0cm

This work was supported by the National Science Council of R.O.C. under
Grant Nos. NSC87-2112-M006-018 and NSC87-2112-M007-040. We also thank the
Center for Theoretical Sciences of the National Science Council of R.O.C.
for partial support.

\newpage
\centerline{\large \bf Figure Captions}
\vskip 0.3cm

\noindent
{\bf Fig. 1.} (a) and (b) Lowest-order diagrams of direct photon production.
(c)-(h) $O(\alpha_s)$ corrections to (a).
\vskip 0.3cm

\noindent
{\bf Fig. 2.} Factorization of direct photon production from hadron
collisions.
\vskip 0.3cm

\noindent
{\bf Fig. 3.} (a) The derivative $p_1^+d{\tilde \phi}/dp_1^+$ in the axial
gauge. (b) General diagrams for the functions $K$ and $G$.
(c) The $O(\alpha_s)$ function $K$. (d) The $O(\alpha_s)$ function $G$. 
\vskip 0.3cm

\noindent
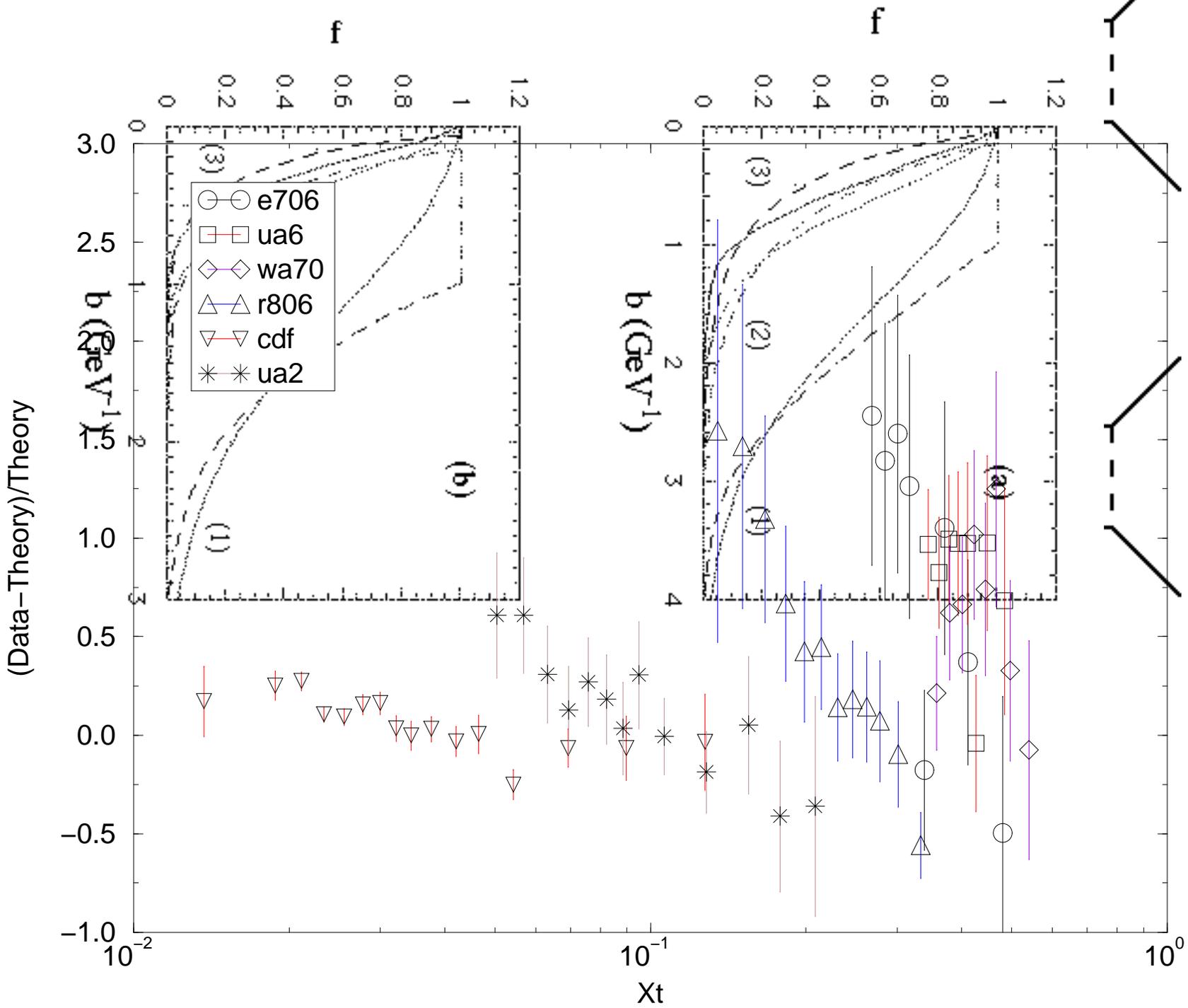
{\bf Fig. 4.} (a) Dependence of $f(\sqrt{s/2},b)$ on $b$ (dashed line) and
its corresponding parametrization (solid line) for (1) $\sqrt{s}=30$ GeV,
(2) $\sqrt{s}=600$ GeV, and (3) $\sqrt{s}=1800$ GeV associated with
(a) the quark distribution function and (b) the gluon distribution
function.
\vskip 0.3cm

\noindent
{\bf Fig. 5.} Compilation of direct photon experiments compared to the NLO
QCD predictions using the CTEQ4M parton distributions
\vskip 0.3cm

\noindent
{\bf Fig. 6.} Compilation of direct photon experiments compared to the
$k_T$-resummed predictions using the CTEQ4M parton distributions

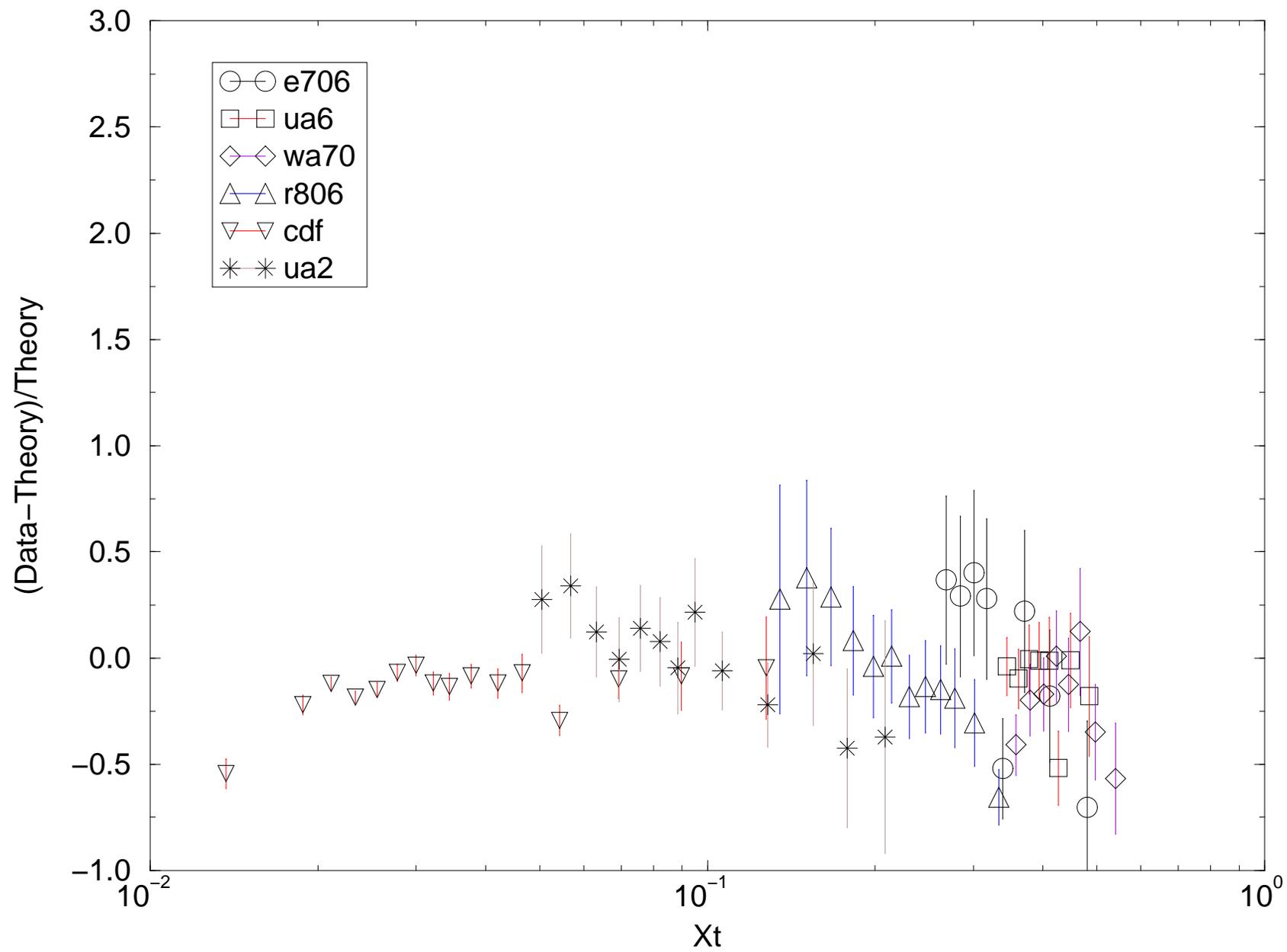
\end{document}